\documentclass[twocolumn,showpacs,showkeys,preprintnumbers,prb,amsmath,amssymb]{revtex4}
\usepackage[dvips]{graphicx}
\usepackage{dcolumn}
\usepackage{bm}
\usepackage{color} 

\newcommand{\sigf}{$\sigma_f(0)$ }
\newcommand{\sigfy}{$\sigma_f(0)$}

\newcommand{\be}{\begin{equation} }

\newcommand{\ene}{\end{equation}}

\newcommand{\eqr}{Eq.~\ref}

\newcommand{\estar}{$E^{*}$ }
\newcommand{\estary}{$E^{*}$}

\newcommand{\figr}{Fig.~\ref}

\newcommand{\hcl}{$H_{c1}$ }

\newcommand{\hcu}{$H_{c2}$ }

\newcommand{\je}{$j(E)$ }

\newcommand{\qp}{quasiparticle }
\newcommand{\qps}{quasiparticles }

\newcommand{\tauee}{$\tau_{ee}$ }

\newcommand{\tauep}{$\tau_{ep}$ }
\newcommand{\tauepy}{$\tau_{ep}$}
\newcommand{\tauey}{$\tau_{\varepsilon}$}
\newcommand{\tc}{$T_{c}$ }

\newcommand{\tcy}{$T_{c}$}

\begin{document}

\preprint{Published in Phys. Rev. B {\bf 82}, 144517 (2010)}
\title{Vortex instability in molybdenum-germanium superconducting films}

\author{Manlai Liang}
\author{Milind N. Kunchur}
\email[Corresponding author email: ]{kunchur@sc.edu}
\homepage{http://www.physics.sc.edu/~kunchur}
\affiliation{Department of Physics and Astronomy, University of South
Carolina, Columbia, SC 29208}

\date{Received 21 August 2010; published 18 October 2010}

\begin{abstract}
We studied the high driving force regime of the current-voltage transport
response in the mixed state of amorphous
molybdenum-germanium superconducting films to the point where the flux
flow becomes unstable.
The observed nonlinear response conforms with the
classic Larkin-Ovchinikov picture with a quasiparticle-energy-relaxation
rate dominated by the quasiparticle recombination process.
The measured energy relaxation rate was found to have a
magnitude and temperature dependence in agreement with theory.
\end{abstract}

\pacs{74.40.Gh, 74.25.Uv, 72.15.Lh, 73.50.Gr, 73.50.Fq}


\keywords{vortices, fluxon, LO}

\maketitle

\section{Introduction and theory}

In a type II superconductor, magnetizing fields between the lower critical
field \hcl and upper critical field \hcu
introduce flux vortices  containing a quantum of
flux $\Phi_{o}=h/2e$. A transport electric current density $j$ perpendicular
to the vortices and to 
exerts a ``Lorentz'' driving force $F_{L} = j \Phi_{o}$.
The consequent vortex motion generates an electric field $E=vB$ and is
opposed by a viscous drag $\eta v$ (where $\eta$ is the coefficient of
viscosity and $v$ the vortex velocity),
so that in steady state $j \Phi_{o} =  \eta v \propto E$ and the response is
Ohmic as long as the flow is not hindered by pinning. This regime of flux motion
corresponds to free flux flow (FFF).
While superficially the physics appears simple,
and bears resemblance to a hydrodynamic system, this resemblance and
apparent simplicity are
deceiving. First of all the so-called ``Lorentz'' force in a superconductor 
actually has the opposite direction to the usual electromagnetic Lorentz
force and
while it has the right magnitude  $F_{L} = j \Phi_{o}$, the derivation of this
 expression is not completely trivial
\cite{degennes,tinkhamtext,narayan,chen,castro,ao}. Second  the hydrodynamic
 analogy leads to the expectation that a narrower vortex would have a lower
 drag coefficient. Under normal circumstances the reality is just the opposite:
dissipation
in and around the core of a vortex arises from generated electric fields acting
upon
 normal quasiparticles leading to Ohmic dissipation \cite{bs} and from irreversible entropy
 transfer occurring between the leading and trailing edges of moving vortices \cite{tinkham}.
 These dissipative processes increase
with the  order-parameter gradient  so that
 $\eta$ is roughly inversely proportional to the vortex core area. Thus under
 normal circumstances  $\eta$ drops as the the vortex expands and hence the
 flux-flow resistance goes up with temperature \cite{unstable,vodolazov}. 

The Ohmic FFF regime discussed above ceases as $j$ and $E$ are increased
to high values  that alter the superconducting state. If the electron-electron
scattering time \tauee is  short compared to the the electron-phonon
scattering time \tauepy, the distribution function of the quasiparticles
becomes thermalized and a finite power dissipation density $jE$ simply
raises the electron temperature
to a value $T'$ above the phonon and substrate temperatures, $T_p$ and $T_0$
respectively. This causes an expansion of the vortex core and a drop in
viscosity leading to a non-monotonic (``N'' shaped) 
\je curve and a consequent
instability. Such a hot-electron instability model was developed 
and experimentally verified by us in
our earlier work \cite{unstable,eprelax}.

A more intricate scenario, proposed by Larkin and Ovchinnikov
\cite{lopaper,lochapter} (LO), arises when \tauee $>$\tauepy .
In this case the quasiparticle distribution function acquires
a non-thermal shape and the vortex dynamics are altered in a
complicated and less obvious way. At high $E$
the quasiparticle population in the
core reduces while it increases outside the core, thus causing the vortex
to shrink while simultaneously reducing the contrast in quasiparticle
density between the two regions. Thus the gradient in the order
parameter $\Delta$
is not boosted to the extent anticipated by the reduced vortex size.
In the meanwhile, the  diminished quasiparticle population in the
core tends to lessen the dissipation and reduce the drag. Thus
overall $\eta$ declines despite the shrinking in size with
increasing $E$. Like the hot-electron case discussed earlier, this
again leads to a non-monotonic \je curve and a vortex instability.
The LO instability has been observed in previous experiments
\cite{Musienko,klein,Doettinger-YBCO,Samoilov,zhili-1996,
Doettinger-MoSi,ruck,babic,armenio,grimaldi2009,grimaldi2010} 
and the combination of heating effects and the LO
mechanism have been considered by various authors
\cite{bezuglyj,gurevich_ciovati,zhili-1998,Zhili-1999,Peroz}.

The relative magnitudes of \tauee and \tauep govern which mechanism
dominates the instability. Standard estimates
\cite{abrikosov_metals} for the scattering times
\tauee=$r\hbar\epsilon_F/k_B^2 T^2$ and \tauep=$r^3\hbar T^2_D/k_B
T^3$ (where $T_D$ is the Debye temperature and $r < 1$ is the phonon
reflection coefficient at the film-substrate interface arising from
acoustic mismatch) give a cross-over temperature of $T_X= r^2k_B
T_D^2/\epsilon_F$. In the present work, we investigated
amorphous molybdenum-germanium (MoGe) films,
which have (estimating from known parameters \cite{carter,graybeal}) 
$T_D$=260 K and $\epsilon_F \approx 10$ eV, and thus have $T_X < 0.6r^2$
K. Since $r<1$,  $T_X$ will be well below 
the temperature range of our experiment (3--6 K). Thus we expect
the non-linear \je response to be dictated  by the LO mechanism,
which is indeed born out by our data.

In the LO theory, the non-linear flux-flow conductivity at high $E$
is given by (Eqs.~38 and 53 of Ref.~\onlinecite{lopaper})
\be \label{LO-sigma} \sigma(E) \approx
\sigma_f \left\{ \frac{1}{1+\left(\frac{E}{E^*}\right)^2} +c \sqrt{1-t}
\right\} \approx
\frac{\sigma_f}{1+\left( \frac{E}{E^*} \right)^2}
, \ene
in terms of the free-flux-flow value $\sigma_f$ of the linear regime
($E\rightarrow 0$ limit in the absence of pinning)
and $E^*$ the critical electric field  at which $j$
attains its maximum value before the \je curve enters a region of negative
differential conductivity.  $t={T}/{T_{c}}$ is the reduced temperature
(\tc is the superconducting transition temperature),
$c$ is an unknown constant of order unity, and
\estar is given by
\be \label{LO-Velocity}
{E^*}^2=\frac{D\sqrt{14\zeta(3)(1-t)}}{\pi\tau_{\varepsilon}}B^2 ,
\ene
where $D$ is the diffusion constant,
$\zeta(x)$ is the Riemann zeta
function, and $\tau_{\varepsilon}$ is the energy relaxation time.

In our previous work \cite{unpinned} we found that the
expressions for \sigf in reference \onlinecite{lochapter} (their Eqs.~22
and 30) did not fit the data well over a significant range.
Instead the following expression \cite{unpinned}
\be
\label{dc-sigma} \sigma_f \approx
\sigma_n + \sigma_n \left(\frac{1-b}{\nu b} \right),
\ene
based on the mean-field result of time-dependent Ginzburg Landau (TDGL)
theory \cite{schmid,caroli,vecris,ullah,dorsey,troy},
more accurately represented the behavior over an extended
range; here
$b =B/\mu_0 H_{c2}$ is the reduced magnetic field (\hcu is the upper
critical magnetizing field) and $\nu \sim 0.3$  is a dimensionless constant.
The
right hand side of \eqr{dc-sigma}  represents a two-fluid-model
sum of the normal conductivity $\sigma_n$ and the flux-flow conductivity
contribution. The
first term is negligible compared to the second term
for the range of conditions where we study the LO nonlinear effect (i.e.,
our mixed-state conductivity is far higher than the normal conductivity)
and also the  $\sigma_n$ term remains constant and is not affected
by $E$ and $B$ fields. Thus combining \eqr{LO-sigma} with the second
term of \eqr{dc-sigma} gives
the following nonlinear \je relationship
\be
\label{LO-IV} j =  E \left[ \left\{
\sigma_n \left(\frac{1-b}{\nu b} \right)
\right\} \left\{ \frac{1}{1+\left({E}/{E^*}\right)^2} \right\} \right].
\ene
Note that this non-linear function is only valid until the vortex stops
shrinking, which occurs at a field \cite{lochapter}
$E_s \sim E^*/(1-t)^{1/4}$. At very high
$E$ the system eventually enters the normal state and then
$\sigma=\sigma_n$.

\section{Experimental techniques}
The samples A, B, and C used in this experiment are exactly the same as
the samples A, B, and C used in our prior work on free flux flow
\cite{unpinned}. The samples
consist of MoGe films of thickness $50$ nm sputtered
onto silicon substrates with 200 nm thick oxide layers using an alloy 
target of atomic composition Mo$_{0.79}$Ge$_{0.21}$. 
The deposition system had a base pressure of $2 \times 10^{-7}$ Torr 
and the argon-gas working pressure
was maintained at 3 mTorr during the sputtering. The
growth rate was 0.15 nm/s.
The samples were
patterned into bridges of length $l=102 \ \mu$m and width $w=6 \
\mu$m using photolithography and argon ion milling.
Some parameters of the samples are as
follows: Sample A: \tcy=5.56 K, $R_n$=555 $\Omega$,
$\mu_0H'_{c2}$=-3.13 T/K and $D = 0.35$ cm$^2$/s. Sample B:
\tcy=5.41 K, $R_n$=555 $\Omega$, $\mu_0H'_{c2}$=-3.13 T/K and $D =
0.35$ cm$^2$/s. Sample C: \tcy=5.01 K, $R_n$=630 $\Omega$,
$\mu_0H'_{c2}$=-3.0 T/K and $D = 0.37$ cm$^2$/s. Here, $R_n$ is the
normal-state resistance, $H'_{c2}  =
d H_{c2}/dT|_{T_c}$ is the upper-critical-field slope, and
the diffusion coefficient D was calculated from
\cite{gorkov1959} $D=-8 k_B/2 \pi e \mu_0 H'_{c2}$.

The cryostat was a Cryomech PT405 pulsed-tube closed-cycle
refrigerator that went down to about 3.2 K. It was fitted inside a
1.3 Tesla GMW 3475-50 water-cooled copper electromagnet. Calibrated
cernox and hall sensors monitored $T$ and $B$ respectively.
The electrical transport measurements were
made with an in-house built pulsed current source, preamplifier
circuitry, and a LeCroy model 9314A digital storage oscilloscope. The pulse
durations are 20 $\mu$s or less, with duty cycles in the few ppm range
to reduce macroscopic heating of the film.
Our previous  review papers \cite{pbreview,mplb} give further details
about the measurement technique and the thermal analysis.

\section{Data and analysis}
\begin{figure}[ht]
\includegraphics[width=0.8\hsize]{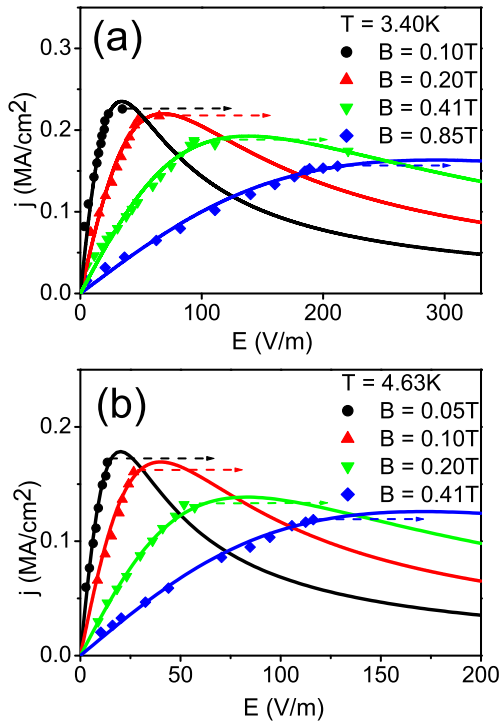}

\vspace{-1em} \caption{ $j$ versus $E$ curves for sample B at two
temperatures (a) 3.40 and (b) 4.63 K, at indicated $B$ values.
The symbols correspond to the measured data and
the solid lines correspond to \eqr{LO-IV}. The dashed arrows indicate
the observed jump in $E$. For the theoretical curves, $\nu$ and $E^*$
were adjusted to the following values: For $T=$3.40 K,
$\nu \approx 0.3$ and $E^*$= 34, 68, 138, and 292 V/m in the order of
ascending $B$. For
$T$=4.63 K,  $\nu \approx 0.2$; and $E^*$= 20, 40, 83, and 171 V/m.}
\label{NegjE}
\end{figure}

\figr{NegjE} shows some examples of nonlinear \je curves. $j$
(and hence the viscous drag force) has a local maximimum value of $j^*$ at
the instability field \estary.
In a current biased circuit, where the source resistance is larger than the
sample resistance as is the case here, $E$ jumps (indicated by arrows)
upon increasing $j$ to the vicinity of the intrinsic $j^*$. In a voltage biased
measurement, where the source resistance is lower than the sample's, the
there will not be a jump in $E$ and instead $j$ will be seen to decrease.
The macroscopically averaged behavior will have a negative $dj/dE$ and
the flux matter fragments into compressional \cite{steps},
shear \cite{shear} or other types of elastic domains such that any
given domain is moving in a response region with $dj/dE>0$ locally.
The macroscopic \je curve will then not follow the primitive curve (e.g.,
\eqr{LO-IV}) but will show steps in the region where $dj/dE<0$.
In the present experiment we are only concerned with the region of the
transport response up to \estar where $dj/dE>0$
macroscopically.

The solid lines in \figr{NegjE} are fits to \eqr{LO-IV} and are seen to
follow the trends of the data. The parameters $\nu$ and \estar (location
of peak) were adjusted to improve the fits, but have fit values of
the expected magnitudes: $\nu \sim 0.3$ and \estar from the peaks is
slightly higher than the position of the actual jumps, which is to be
expected and has been observed by others (e.g., reference
\onlinecite{babic}). For subsequent analysis, we take \estar to be the
actual measured value of $E$ at the threshold of the jump.

\begin{figure}[ht]
\includegraphics[width=0.75\hsize]{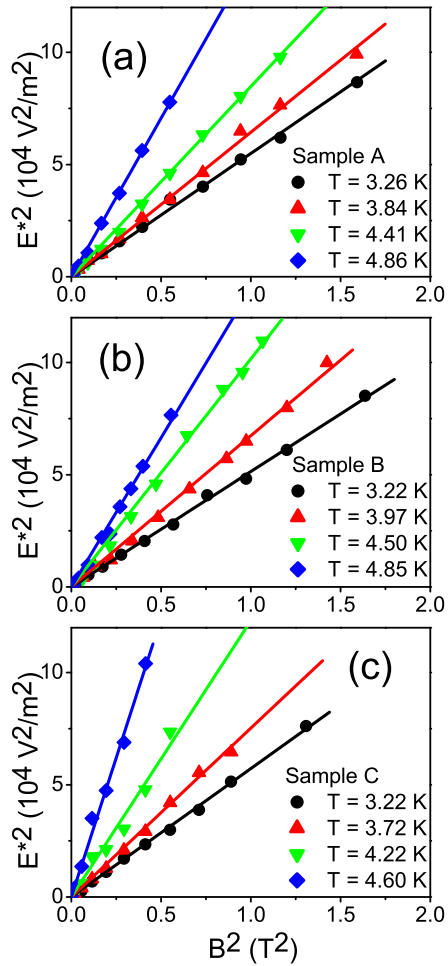}
\vspace{-1em} \caption{${E^*}^2$ versus $B^2 $ for samples A, B, and C
at indicated temperatures showing proportionality between the two
quantities as per \eqr{LO-Velocity}.
The lines are least-squares linear fits to the data (symbols).}
\label{GVABC}
\end{figure}

\begin{figure}[ht]
\includegraphics[width=0.75\hsize]{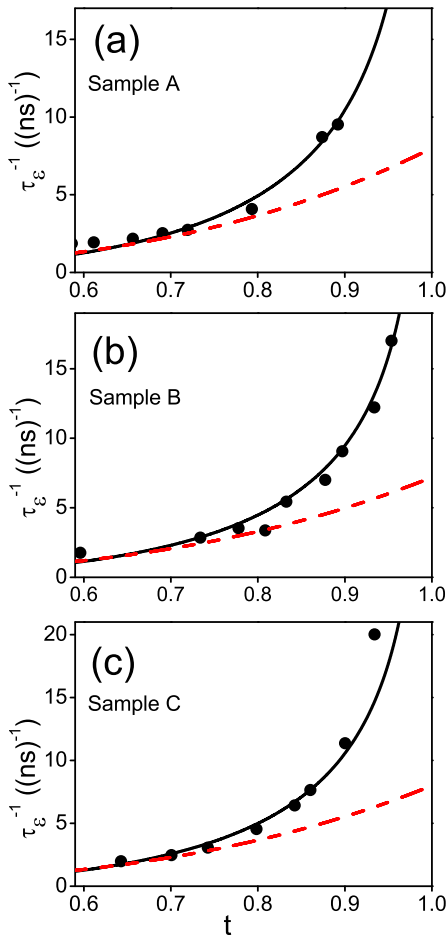}
\vspace{-1em} \caption{Temperature dependence of the energy-relaxation
rate. The symbols represent the experimental values deduced from
\eqr{LO-Velocity} and the measured \estary.
The solid black lines are fits to \eqr{taur}
 (quasiparticle recombination process) with the
values  ${\cal{T}}=3.4 \times10^{-11}$ s, $3.7\times10^{-11}$ s, and
$3.3\times10^{-11}$ for samples A--C respectively. The dashed red lines
are fits to $\tau_{\varepsilon}^{-1}=
a t^{7/2}$ (expected for inelastic quasiparticle-phonon
scattering \cite{kaplan}) with $a$ = $8\times 10^{9}$ s$^{-1}$,
$7 \times 10^{9}$ s$^{-1}$, and $8\times 10^{9}$ s$^{-1}$
for samples A--C respectively.}
\label{Rate}
\end{figure}

\figr{GVABC} plots \estary$^2$, obtained from the \je curves as discussed
above, against $B^2$. In agreement with \eqr{LO-Velocity}, the two
quantities are directly proportional to each other (i.e., the critical vortex
velocity $v^*$=\estary /$B$ is independent of $B$). From the measured slope and
\eqr{LO-Velocity}, we obtain the energy relaxation time \tauey .
\figr{Rate} plots the corresponding relaxation rate
$\tau_{\varepsilon}^{-1}$ against the reduced temperature
for each of the three samples.

As discussed in the introduction, the LO effect occurs when
\tauee is long compared with \tauepy, resulting in a non-thermal shape
of the quasiparticle distribution function. The extent of the distribution
function distortion is controlled by the rate of
energy relaxation from quasiparticles to phonons, which occurs through two
processes: one process is the
inelastic scattering between a \qp and a phonon and the other is the
recombination of two \qps to form a Cooper pair with the emission of a phonon.
As discussed by Kaplan et al. \cite{kaplan},
the energy relaxation is mainly dominated by the latter recombination process
which has a rate that can be written as:
\be
\label{taur} \tau^{-1}_{\varepsilon} = {\cal{T}}^{-1}
{\left( \frac{k_B
T}{\Delta}\right)}^{1/2} \exp\left(-\frac{\Delta}{k_B T}\right), \ene
where ${\cal{T}}$ is a temperature independent characteristic time
constant (in the terminology of Kaplan et al. \cite{kaplan},
${\cal{T}} \approx \tau_0 /55$) and $\Delta$ is the temperature dependent
superconducting energy gap. Taking the BCS (Bardeen-Cooper-Schrieffer)
temperature dependence for $\Delta$ we are able to fit the data
in \figr{Rate} with \eqr{taur} (solid lines) with
${\cal{T}} \approx 3.5\times 10^{-11}$ s. As can be seen, the measured
functional form of \tauey$^{-1}(t)$ is in agreement with \eqr{taur}.
While there is insufficient
information in the literature to theoretically compute the magnitude of
${\cal{T}}$ for comparison, Table~I of
Reference~\onlinecite{kaplan} lists values of $\tau_0$ ($\approx 55
\times {\cal{T}}$) for various other materials.
Although there is an enormous range in ${\cal{T}}$---from
$8 \times 10^{-13}$ to $4 \times 10^{-5}$ s for various materials---it
is interesting to
observe that tantalum with about the same \tc (4.5 K) and $T_D$
(240 K) as MoGe (for which \tcy $\approx$5.3 K and $T_D$$\approx$260 K)
has a value of  ${\cal{T}}$ ($3.3\times 10^{-11}$ s) that is comparable
to the one we obtained for MoGe ($3.5 \times 10^{-11}$ s).
(\tc and $T_D$ are parameters that are indicative of the
electron-phonon coupling and phonon density of states respectively.)
If quasiparticle-phonon scattering is the dominant
relaxation process, rather than quasiparticle recombination, then the rate is
given by a function of the form \cite{kaplan} $\tau_{\varepsilon}^{-1} 
\propto t^{7/2}$
instead of \eqr{taur}. This power-law function (taken with an adjustable
constant of proportionality $a$) 
is plotted as dashed red lines on \figr{Rate}
and is clearly at odds with the data.
Thus our study of the vortex instability is able to distinguish the two
routes of energy decay and
provides a confirmation of the recombination rate expression of \eqr{taur}.

\section{Conclusions}
In conclusion, we have studied the high driving force regime of vortex
dynamics in one of the simplest and nearly model
superconductors (unpinned, isotropic, low-temperature,
weak-coupling BCS, etc.). In recent work \cite{unpinned} we
found that for the FFF regime these MoGe films provided a detailed
confirmation of the TDGL mean-field prediction for \sigfy ,
while the LO expressions for the
same regime showed very limited applicability. On the other hand in
the present work we find that the LO expression  (\eqr{LO-sigma})
for the nonlinear modulation factor for
\sigf and the LO result  (\eqr{LO-Velocity}) for the relationship  between
the instability electric field \estar and
the energy relaxation time \tauey ,
are well obeyed in this system.
Furthermore we were able to distinguish between the two principal
quasiparticle-phonon energy relaxation processes and
confirm the predicted temperature
dependence (\eqr{taur}) for the recombination process.
Both regimes of instability, the hot-electron as well as
the distribution-function type, thus provide a valuable tool for
investigating key scattering processes through the response of the mixed
state.

\section{Acknowledgements}
The authors gratefully acknowledge 
Jiong Hua, Zhili Xiao, James M. Knight, and Boris I. Ivlev.
This work was supported by the U. S. Department of Energy through grant
number DE-FG02-99ER45763.

\end{document}